\documentclass[12pt,a4j]{article}
\usepackage[dvips]{graphicx}
\usepackage{enumerate}
\usepackage{amsmath}
\usepackage{amsfonts}

\begin{document}
\baselineskip=7mm
\centerline{\bf The $N$-soliton solution of a two-component}\par
\centerline{\bf modified nonlinear Schr\"odinger equation  }\par
\bigskip
\centerline{Yoshimasa Matsuno\footnote{{\it E-mail address}: matsuno@yamaguchi-u.ac.jp}}\par

\centerline{\it Division of Applied Mathematical Science,}\par
\centerline{\it Graduate School of Science and Engineering} \par
\centerline{\it Yamaguchi University, Ube, Yamaguchi 755-8611, Japan} \par
\bigskip
\bigskip
\leftline{\bf ABSTRACT}\par
The $N$-soliton solution is presented for a two-component modified nonlinear Schr\"odinger equation which describes the propagation of short pulses in
birefringent optical fibers.  The solution is found to be expressed in terms of determinants. The proof
of the solution is carried out by means of an elementary theory of determinants. The generalization of the 2-component system to the 
multi-component system is discussed as well as a (2+1)-dimensional  nonlocal equation arising from its continuum limit. \par
 \par

\newpage
\leftline{\bf  1. Introduction} \par
We consider the following two-component system of nonlinear evolution equations which is a hybrid of the
coupled nonlinear Schr\"odinger (NLS) equation and coupled derivative NLS equation
$${\rm i}\,q_{j,t}+q_{j,xx}+\mu\left(\sum_{k=1}^2|q_k|^2\right)q_j+{\rm i}\gamma\left[\left(\sum_{k=1}^2|q_k|^2\right)q_j \right]_x=0,\quad (j=1, 2), \eqno(1.1)$$
where $q_j=q_j(x,t) \ (j=1, 2)$ are the slowly varying complex envelopes for polarizations, $\mu$ and $\gamma$ are real constants
and subscripts $x$ and $t$ appended to $q_j$ denote  partial differentiations. This system of equations has been derived as a model for describing
the propagation of short pulses in birefringent optical fibers [1]. Its integrability has been established by constructing the Lax pair and an
infinite number of conservation laws [2]. Hence, the initial value problem can be formulated in principle by means of the inverse scattering method. 
A few exact solutions are now available for the system (1.1). Actually,  one- and two-soliton solutions have been
presented  by employing Hirota's bilinear transformation method [3, 4]. The $N$-soliton solution of the system (1.1) with $\mu=0$ was also obtained quite recently
by means of the Darboux transformation where $N$ is an arbitrary positive integer [5].  \par
The purpose of this Letter is to construct the bright  $N$-soliton solution of the system (1.1) within the framework of the bilinear formalism. We present  a compact determinantal
expression of the $N$-soliton solution which would be suitable for investigating the interaction process of solitons. The proof of the solution is performed by using an elementary theory of 
determinants. In concluding remarks, the generalization of the system (1.1) to the $n$-component system $(n \geq 3)$ is briefly discussed as well as 
a (2+1)-dimensional nonlocal modified NLS equation arising from its continuum limit. Furthermore, some issues associated with the system (1.1) will be addressed. \par
\bigskip
\leftline{\bf 2. Bilinear form and $N$-soliton solution}\par
\leftline{\it 2.1. Bilinear form}\par
We first apply the gauge transformations 
$$q_j=u_j\,{\rm exp}\left[-{{\rm i\gamma}\over 2}\int_{-\infty}^x\sum_{k=1}^2|u_k|^2dx\right],\quad (j=1, 2), \eqno(2.1)$$
to the system (1.1) and transform it to the system of equations for $u_j$
$${\rm i}\,u_{j,t}+u_{j,xx}+\mu\left(\sum_{k=1}^2|u_k|^2\right)u_j+{\rm i}\gamma\left(\sum_{k=1}^2u_k^*u_{k,x}\right)u_j=0,\quad (j=1, 2), \eqno(2.2)$$
where the asterisk denotes complex conjugate.
Note that we have imposed the boundary conditions $q_j\rightarrow 0, u_j\rightarrow 0$ as $|x|\rightarrow \infty$ to give the bright soliton solutions.
By means of the dependent variable transformations
$$u_j={g_j\over f},\quad (j=1, 2), \eqno(2.3)$$
the system (2.2) can be transformed to the following system of bilinear equations for $f$ and $g_j$ [4]
$$({\rm i}D_t+D_x^2)g_j\cdot f=0, \quad (j=1, 2), \eqno(2.4)$$
$$D_xf\cdot f^*={{\rm i}\gamma\over 2}\sum_{k=1}^2|g_k|^2, \eqno(2.5)$$
$$D_x^2f\cdot f^*=\mu\sum_{k=1}^2|g_k|^2+{{\rm i}\gamma\over 2}\sum_{k=1}^2D_xg_k\cdot g_k^*. \eqno(2.6)$$
Here, the bilinear operators $D_x$ and $D_t$ are defined by
$$D_x^mD_t^nf\cdot g=\left({\partial\over\partial x}-{\partial\over\partial x^\prime}\right)^m
\left({\partial\over\partial t}-{\partial\over\partial t^\prime}\right)^n
f(x, t)g(x^\prime,t^\prime)\Big|_{ x^\prime=x,\,t^\prime=t},  \eqno(2.7)$$
where $m$ and $n$ are nonnegative integers.
It follows from (2.3) and (2.5) that
$$-{{\rm i}\gamma\over 2}\sum_{k=1}^2|u_k|^2={\partial\over\partial x}\,{\rm ln}\,{f^*\over f}, \eqno(2.8)$$
which, substituted into (2.1), yields the solution of the system (1.1) in the form $q_j=g_jf^*/f^2\ (j=1, 2)$.
\par
\medskip
\leftline{\it  2.2. N-soliton solution} \par
We now state the main result in this Letter:\par
 {\it The $N$-soliton solution of the system of bilinear equations (2.4), (2.5) and (2.6) is given by the
 determinants $f$ and $g_j\ (j=1, 2)$ where
$$f=\begin{vmatrix} A & I\\ -I & B\end{vmatrix}, \quad g_j=\begin{vmatrix} A &I & {\bf z}^T \\ -I & B &{\bf 0}^T\\
                                                                            {\bf 0} & -{\bf a}_j^*
                                                                            & 0 \end{vmatrix},\quad (j=1, 2). \eqno(2.9)$$
 Here, $A, B$ and $I$ are $N\times N$ matrices and ${\bf z}, {\bf a}_j$ and ${\bf 0}$ are $N$-component row vectors defined below and the symbol $T$ 
 denotes the transpose:
 $$A=(a_{jk})_{1\leq j,k\leq N}, \quad a_{jk}={1\over 2}\,{z_jz_k^*\over p_j+p_k^*}, \quad z_j={\rm exp}(p_jx+{\rm i}p_j^2t), \eqno(2.10a)$$
 $$B=(b_{jk})_{1\leq j,k\leq N}, \quad b_{jk}={(\mu+{\rm i}\gamma p_k)c_{jk}\over p_j^*+p_k}, \quad c_{jk}=\sum_{s=1}^2\alpha_{sj}\alpha^*_{sk},\eqno(2.10b)$$ 
$$I=(\delta_{jk})_{1\leq j,k\leq N}, : N\times N\ {\it unit\  matrix}, \eqno(2.10c)$$
$${\bf z}=(z_1, z_2, ..., z_N), \quad  {\bf a}_j=(\alpha_{j1}, \alpha_{j2}, ..., \alpha_{jN}), \quad
{\bf 0}=(0, 0, ..., 0). \eqno(2.10d)$$ } 
The above $N$-soliton solution involves  $3N$ complex parameters $p_j$ and $\alpha_{1j}, \alpha_{2j}\ (j=1, 2, ..., N).$
The former parameters determine the amplitude and velocity of the solitons whereas the latter ones determine the polarizations and the 
envelope phases of the solitons. In the special case of $N=1,2$, (2.9) and (2.10) reproduce the 1- and 2-soliton solutions presented in [4].
Note, however that the compact determinantal expression of the solution has been obtained here for the first time. It is also remarked that
in the single-component case (i.e., the system
 (1.1) with $q_2=0$), our solutions reduce to the soliton solutions given in [6].
 \par
 \medskip
\leftline{\it 2.3. Remark} \par
If we introduce the transformations
$$f=\tilde f, \quad g_j={\rm exp}\left[{\rm i}\left\{{\mu\over\gamma}\,\tilde x+\left({\mu\over\gamma}\right)^2\,\tilde t\right\}\right]\tilde g_j, \quad (j=1, 2), \eqno(2.11a)$$
$$x=\tilde x+{2\mu\over\gamma}\,\tilde t, \quad t=\tilde t, \eqno(2.11b)$$
then the bilinear equations (2.4), (2.5) and (2.6) recast to
$$({\rm i}D_{\tilde t}+D_{\tilde x}^2)\tilde g_j\cdot \tilde f=0, \quad (j=1, 2), \eqno(2.12)$$
$$D_{\tilde x}\tilde f\cdot \tilde f^*={{\rm i}\gamma\over 2}\sum_{k=1}^2|\tilde g_k|^2, \eqno(2.13)$$
$$D_{\tilde x}^2\tilde f\cdot \tilde f^*={{\rm i}\gamma\over 2}\sum_{k=1}^2D_{\tilde x}\tilde g_k\cdot \tilde g_k^*, \eqno(2.14)$$
respectively. Thus, the form of Eqs. (2.4) and (2.5) is invariant whereas (2.6) becomes a simplified equation with $\mu=0$. 
Consequently, the proof of the $N$-soliton solution may be performed for the corresponding solution with $\mu=0$. 
This reduces the total amount of calculations considerably, in particular for Eq. (2.6). Hence, in the analysis developed in Sec. 3, we
put $\mu=0$ without loss of generality.
\par
\bigskip
\leftline{\bf 3. Notation and formulas for determinants}\par
To perform the proof of the $N$-soliton solution  concisely, we introduce some notations and write the basic formulas for
determinants.  \par
\leftline{\it 3.1. Notation}\par
We define the following matrices associated with the $N$-soliton solution (2.9):
$$D=\begin{pmatrix} A & I\\ -I & B\end{pmatrix},\eqno(3.1)$$
$$D({\bf a};{\bf b})=\begin{pmatrix} A &I & {\bf 0}^T \\ -I & B &{\bf b}^T\\
                                                                            {\bf 0} & {\bf a}& 0 \end{pmatrix}, \eqno(3.2)$$
$$D({\bf a};{\bf z})=\begin{pmatrix} A &I & {\bf z}^T \\ -I & B &{\bf 0}^T\\
                                                                            {\bf 0} & {\bf a}& 0 \end{pmatrix}, \eqno(3.3)$$
$$D({\bf z}^*;{\bf z})=\begin{pmatrix} A &I & {\bf z}^T \\ -I & B &{\bf 0}^T\\
                                                                            {\bf z}^* & {\bf 0}& 0 \end{pmatrix}. \eqno(3.4)$$
Note the position of the vectors ${\bf a}$, ${\bf b}, {\bf z}$ and ${\bf z}^*$ in the above expressions where ${\bf a}$ and ${\bf b}$
represent ${\bf a}_j^*$ and ${\bf a}_j\ (j=1, 2)$, respectively. The matrices which include more than two vectors
will be introduced as well. \par
\medskip
\leftline{\it 3.2. Formulas for determinants}\par
In the process of the proof, we use some basic formulas  in addition to the fundamental properties of determinants, which we shall now write
down for later convenience.
Let $A=(a_{jk})_{1\leq j,k\leq M}$ be an $M\times M$ matrix with $M$ being an arbitrary 
positive integer and $A_{jk}$ be the cofactor of the element $a_{jk}$.
Then, we have [7]
$${\partial\over\partial x}|A|=\sum_{j,k=1}^M{\partial a_{jk}\over\partial x}A_{jk}, \eqno(3.5)$$
$$\begin{vmatrix} A & {\bf a}^T\\ {\bf b} & z\end{vmatrix}=|A|z-\sum_{j,k=1}^MA_{jk}a_jb_k,  \eqno(3.6)$$
$$|A({\bf a}_1, {\bf a}_2; {\bf b}_1, {\bf b}_2)||A|= |A({\bf a}_1; {\bf b}_1)||A({\bf a}_2; {\bf b}_2)|-|A({\bf a}_1; {\bf b}_2)||A({\bf a}_2; {\bf b}_1)|. \eqno(3.7)$$
The formula (3.5) is the differentiation rule of the determinant and (3.6) is the expansion formula for a bordered determinant
with respect to the last row and column.
The formula (3.7) is Jacobi's identity and it will play a central role in the proof. \par
The following two formulas are also employed frequently: 
$$|A({\bf a}_1,..., {\bf a}_n; {\bf b}_1,..., {\bf b}_n)||A|^{n-1}=\begin{vmatrix} |A({\bf a}_1;{\bf b}_1)| &\cdots & |A({\bf a}_1;{\bf b}_n)| 
                                                                                    \\ \vdots & \ddots & \vdots                                                                                                                                                                                                                                                       
                                                                                    \\ |A({\bf a}_n;{\bf b}_1)| &\cdots & |A({\bf a}_n;{\bf b}_n)|
                                                                                     \end{vmatrix},\quad (n \geq 2), \eqno(3.8)$$
$$|A+\sum_{s=1}^2{\bf b}_s^T{\bf a}_s|=|A|-\sum_{s=1}^2|A({\bf a}_s;{\bf b}_s)|+|A({\bf a}_1,{\bf a}_2; {\bf b}_1,{\bf b}_2)|. \eqno(3.9)$$
The formula (3.8) can be proved by a mathematical induction. The case $n=2$ reduces to Jacobi's identity (3.7).  
The notation ${\bf b}_s^T{\bf a}_s$ on the left-hand side of (3.9) represents an $M\times M$ matrix whose $(j,k)$ element is given by $b_{sj}a_{sk}$.
The formula (3.9) is derived simply using the property of the bordered determinant. \par
\bigskip
\leftline{\bf 4. Proof of the $N$-soliton solution}\par
In this section, we perform the proof of the $N$-soliton solution by means of an elementary theory of determinants.
To this end, we need the differentiation rules of the determinants $f$ and $g_j$ with respect to $x$ and $t$ as well as 
their complex conjugate expressions. We first write down them and then proceed to the proof. \par
\medskip
\leftline{\it 4.1. Differentiation rules of $f$ and $g_j$}\par
In terms of the notation introduced in Sec. 3.1, we have
$$f=|D|, \quad g_j=-|D({\bf a}_j^*;{\bf z})|. \eqno(4.1)$$
The differentiation rules for these determinants are derived simply with use of the formulas (3.5) and (3.6). They read
$$f_t=-{{\rm i}\over 2}\left\{|D({\bf z}^*;{\bf z}_x)|-|D({\bf z}_x^*;{\bf z})|\right\}, \eqno(4.2)$$
$$f_x=-{1\over 2}|D({\bf z}^*;{\bf z})|, \eqno(4.3)$$
$$f_{xx}=-{1\over 2}\left\{|D({\bf z}^*;{\bf z}_x)|+|D({\bf z}_x^*;{\bf z})|\right\}, \eqno(4.4)$$
$$g_{j,t}=-|D({\bf a}_j^*;{\bf z}_t)|+{{\rm i}\over 2}|D({\bf a}_j^*,{\bf z}^*;{\bf z},{\bf z}_x)|, \eqno(4.5)$$
$$g_{j,x}=-|D({\bf a}_j^*;{\bf z}_x)|, \eqno(4.6)$$
$$g_{j,xx}=-|D({\bf a}_j^*;{\bf z}_{xx})|+{1\over 2}|D({\bf a}_j^*,{\bf z}^*;{\bf z}_x,{\bf z})|. \eqno(4.7)$$
For an instructive purpose, we give a proof of (4.3). The other formulas can be proved in the same way.
First, let $D=(d_{jk})_{1\leq j,k\leq 2N}$ and $D_{jk}$ be the cofactor of $d_{jk}$. We then apply the formula (3.5)
to the determinant $f$ given in (2.9) to obtain $f_x={1\over 2}\sum_{j,k=1}^{N}D_{jk}z_jz_k^*$.
This can be put into the form (4.3) in view of the formula (3.6) with $z=0$ and the definition (3.4).
The key in the proof is that the factor $(p_j+p_k^*)^{-1}$  in the element  $a_{jk}$ of the matrix $A$ has been cancelled after
differentiation with respect to $x$. \par
We compute the complex conjugate expressions of these differentiations. To this end, we note from (2.10) that $A^*=A^T$ and $B^*=B^T-{\rm i}\gamma C^T$ where
$C$ is an $N\times N$ matrix with elements $c_{jk}$  defined by $(2.10b)$. These relations lead to the expression of $f^*$
$$f^*=\begin{vmatrix} A & I\\ -I & B-{\rm i}\gamma C\end{vmatrix}. \eqno(4.8)$$
Using the formula (3.9), we can expand $f^*$ in powers of $\gamma$
$$f^*=|D|+{\rm i}\gamma \sum_{j=1}^2|D({\bf a}_j^*;{\bf a}_j)|-\gamma^2|D({\bf a}_1^*,{\bf a}_2^*;{\bf a}_1,{\bf a}_2)|. \eqno(4.9)$$
The similar procedure applied to  $f_x^*$ and $g_j^*$ gives
$$f_x^*=-{1\over 2}|D({\bf z}^*;{\bf z})|-{{\rm i}\gamma\over 2} \sum_{j=1}^2|D({\bf z}^*,{\bf a}_j^*;{\bf z},{\bf a}_j)|
+{\gamma^2\over 2}|D({\bf z}^*,{\bf a}_1^*,{\bf a}_2^*;{\bf z},{\bf a}_1,{\bf a}_2)|, \eqno(4.10)$$
$$g_j^*=|D({\bf z}^*;{\bf a}_j)|+{\rm i}\gamma\sum_{k=1}^2|D({\bf z}^*,{\bf a}_k^*;{\bf a}_j,{\bf a}_k)|. \eqno(4.11)$$
\par
The above formulas will be used to prove Eqs. (2.4) and (2.5).  For the proof of Eq. (2.6), on the other hand, we need another differentiation rules, which we shall now describe.
First, we rewrite $f$ in the form
$$f=\begin{vmatrix} \tilde A & I\\ -I & \tilde B\end{vmatrix}, \eqno(4.12a)$$
where $\tilde A$ and $\tilde B$ are $N\times N$ matrices defined by
$$\tilde A=(\tilde a_{jk})_{1\leq j,k\leq N}, \quad \tilde a_{jk}={1\over 2}\,{1\over p_j+p_k^*},  \eqno(4.12b)$$
 $$\tilde B=(\tilde b_{jk})_{1\leq j,k\leq N}, \quad \tilde b_{jk}={{\rm i}\gamma c_{jk}p_k\over p_j^*+p_k}\,z_j^*z_k.\eqno(4.12c)$$ 
Using the formulas (3.5) and  (3.6), we obtain an alternative expression of $f_x$
$$f_x=-{\rm i}\gamma\sum_{j=1}^2|D({\bf b}_j^*;{\bf a}_j)|, \eqno(4.13a)$$
where ${\bf b}_j$ is an $N$-component row vector given by
$${\bf b}_j=(\alpha_{j1}p_1^*, \alpha_{j2}p_2^*,...,\alpha_{jN}p_N^*). \eqno(4.13b)$$
Differentiation of $(4.13a)$ with respect to $x$ gives
$$f_{xx}={{\rm i}\gamma\over 2}\sum_{j=1}^2|D({\bf z}^*,{\bf b}_j^*;{\bf z},{\bf a}_j)|, \eqno(4.14)$$
Similarly, we have
$$g_{j,x}=-|D({\bf b}_j^*;{\bf z})|+{\rm i}\gamma\sum_{k=1}^2|D({\bf a}_j^*,{\bf b}_k^*;{\bf z},{\bf a}_k)|. \eqno(4.15)$$
The complex conjugate expressions of $f_x$, $f_{xx}$ and $g_{j,x}$ are derived by  referring to the procedure used to obtain $f^*$.
They read
$$f_x^*={\rm i}\gamma\sum_{j=1}^2|D({\bf a}_j^*;{\bf b}_j)|
-\gamma^2\sum_{j,k=1}^2|D({\bf a}_j^*,{\bf a}_k^*;{\bf b}_j,{\bf a}_k)|, \eqno(4.16)$$
$$f_{xx}^*=-{{\rm i}\gamma\over 2}\sum_{j=1}^2|D({\bf z}^*,{\bf a}_j^*;{\bf z},{\bf b}_j)|
+{\gamma^2\over 2}\sum_{j,k=1}^2|D({\bf z}^*,{\bf a}_j^*,{\bf a}_k^*;{\bf z},{\bf b}_j,{\bf a}_k)|, \eqno(4.17)$$
$$g_{j,x}^*=|D({\bf z}^*;{\bf b}_j)|+{\rm i}\gamma \sum_{k=1}^2\Big\{|D({\bf z}^*,{\bf a}_k^*;{\bf b}_j,{\bf a}_k)|+|D({\bf z}^*,{\bf a}_k^*;{\bf a}_j,{\bf b}_k)|\Big\}$$
$$-\gamma^2\Big\{|D({\bf z}^*,{\bf a}_1^*,{\bf a}_2^*;{\bf b}_j,{\bf a}_1,{\bf a}_2)|
+\sum_{k,l=1}^2|D({\bf z}^*,{\bf a}_k^*,{\bf a}_l^*;{\bf a}_j,{\bf b}_k,{\bf a}_l)|\Big\}. \eqno(4.18)$$
It follows from (4.3) and $(4.13a)$ that
$$|D({\bf z}^*;{\bf z})|=2{\rm i}\gamma\sum_{j=1}^2|D({\bf b}_j^*;{\bf a}_j)|. \eqno(4.19)$$
Equating (4.10) and (4.16) and using (4.19) to eliminate $|D({\bf z}^*;{\bf z})|$, we find that
$$\sum_{j=1}^2\left\{|D({\bf a}_j^*;{\bf b}_j)|+|D({\bf b}_j^*;{\bf a}_j)|\right\}=-{1\over 2}\sum_{j=1}^2|D({\bf z}^*,{\bf a}_j^*;{\bf z},{\bf a}_j)|$$
$$-{\rm i}\gamma \Big\{\sum_{j,k=1}^2|D({\bf a}_j^*,{\bf a}_k^*;{\bf b}_j,{\bf a}_k)|+{1\over 2}|D({\bf z}^*,{\bf a}_1^*,{\bf a}_2^*;{\bf z},{\bf a}_1,{\bf a}_2)|\Big\}. \eqno(4.20)$$
\medskip
\leftline{\it 4.2. Proof of Eq. (2.4)}\par
Let $P_1$ be
$$P_1=({\rm i}D_t+D_x^2)g_j\cdot f. \eqno(4.21)$$
Substituting (4.1)-(4.7) into (4.21), $P_1$ becomes
$$P_1=-|D({\bf a}_j^*,{\bf z}^*;{\bf z},{\bf z}_x)||D|+|D({\bf a}_j^*;{\bf z})||D({\bf z}^*;{\bf z}_x)|-|D({\bf a}_j^*;{\bf z}_x)||D({\bf z}^*;{\bf z})|$$
$$-\left\{{\rm i}|D({\bf a}_j^*;{\bf z}_t)|+|D({\bf a}_j^*;{\bf z}_{xx})|\right\}. \eqno(4.22)$$
If we use Jacobi's identity (3.7) and the fundamental formula 
$\alpha|D({\bf a};{\bf b}_1)|+\beta|D({\bf a};{\bf b}_2)|=|D({\bf a};\alpha{\bf b}_1+\beta{\bf b}_2)|$ \ $(\alpha, \beta \in \mathbb{C})$, $P_1$ simplifies to
$$P_1=-|D({\bf a}_j^*;{\rm i}{\bf z}_t+{\bf z}_{xx})|. \eqno(4.23)$$
Since ${\rm i}\,{\bf z}_t+{\bf z}_{xx}={\bf 0}$ by $(2.10a)$, the last column of the  determinant consists only of zero elements. Hence, we see that $P_1=0$, which completes the proof of Eq. (2.4). \par
\medskip
\leftline{\it 4.3. Proof of Eq. (2.5)}\par
Let $P_2$ be
$$P_2=D_xf\cdot f^*-{{\rm i}\gamma\over 2}\sum_{k=1}^2|g_k|^2. \eqno(4.24)$$
If we substitute (4.1), (4.3) and (4.9)-(4.11) into (4.24) and use Jacobi's identity (3.7), we can reduce $P_2$ to
$$P_2={\gamma^2\over 2}\Big\{-|D({\bf z}^*,{\bf a}_1^*,{\bf a}_2^*;{\bf z},{\bf a}_1,{\bf a}_2)||D|+|D({\bf z}^*;{\bf z})||D({\bf a}_1^*,{\bf a}_2^*;{\bf a}_1,{\bf a}_2)|$$
$$-|D({\bf a}_1^*;{\bf z})||D({\bf z}^*,{\bf a}_2^*;{\bf a}_1,{\bf a}_2)|+|D({\bf a}_2^*;{\bf z})||D({\bf z}^*,{\bf a}_1^*;{\bf a}_1,{\bf a}_2)|\Big\}. \eqno(4.25)$$
Assume that the matrix $D$ is non-singular, i.e., $|D|\not=0$ which ensures the regularity of $u_j$ from (2.3). Multiplying $|D|$ on both sides of (4.25) and using Jacobi's identity (3.7)
as well as the formula (3.8) with $n=3$, we see that $|D|P_2=0$, implying that $P_2=0$. This completes the proof of Eq. (2.5). \par
\medskip
\leftline{\it 4.4. Proof of Eq. (2.6)}\par
The proof of Eq. (2.6) is very complicated compared with that of Eq. (2.4) and (2.5) although it can be performed straightforwardly.
Let $P_3$ be
$$P_3=D_x^2f\cdot f^*-{{\rm i}\gamma\over 2}\sum_{k=1}^2D_xg_k\cdot g_k^*. \eqno(4.26)$$
By using (4.1) and (4.13)-(4.18) and applying Jacobi's identity (3.7) and the formula (3.8) with $n=3$, $P_3$ can be written, after some lengthy calculations, in the form
$$P_3={{\rm i}\gamma\over 2}P_{31}+{\gamma^2\over 2}P_{32}+{{\rm i}\gamma^3\over 2}P_{33}, \eqno(4.27a)$$
with
$$P_{31}= |D({\bf z}^*;{\bf z})|\sum_{j=1}^2\Big\{|D({\bf a}_j^*;{\bf b}_j)|+|D({\bf b}_j^*;{\bf a}_j)|\Big\}, \eqno(4.27b)$$
$$P_{32}=-\sum_{j,k=1}^2\Big\{|D({\bf b}_j^*;{\bf a}_j)||D({\bf z}^*,{\bf a}_k^*;{\bf z},{\bf a}_k)|+|D({\bf z}^*;{\bf z})||D({\bf a}_j^*,{\bf a}_k^*;{\bf b}_j,{\bf a}_k)|\Big\}, \eqno(4.27c)$$
$$P_{33}=\sum_{j=1}^2\Big\{-|D({\bf a}_1^*,{\bf a}_2^*;{\bf a}_1,{\bf a}_2)||D({\bf z}^*,{\bf b}_j^*;{\bf z},{\bf a}_j)|
-|D({\bf z}^*,{\bf a}_1^*;{\bf a}_1,{\bf a}_2)||D({\bf a}_2^*,{\bf b}_j^*;{\bf z},{\bf a}_j)|$$
$$+|D({\bf z}^*,{\bf a}_2^*;{\bf a}_1,{\bf a}_2)||D({\bf a}_1^*,{\bf b}_j^*;{\bf z},{\bf a}_j)|\Big\}. \eqno(4.27d)$$
If we introduce (4.19) and (4.20) into $P_{31}$, $P_3$ becomes
$$P_3={{\rm i}\gamma^3\over 2}\sum_{j=1}^2\Big\{|D({\bf b}_j^*;{\bf a}_j)||D({\bf z}^*,{\bf a}_1^*,{\bf a}_2^*;{\bf z},{\bf a}_1,{\bf a}_2)|
-|D({\bf a}_1^*,{\bf a}_2^*;{\bf a}_1,{\bf a}_2)||D({\bf z}^*,{\bf b}_j^*;{\bf z},{\bf a}_j)|$$
$$-|D({\bf z}^*,{\bf a}_1^*;{\bf a}_1,{\bf a}_2)||D({\bf a}_2^*,{\bf b}_j^*;{\bf z},{\bf a}_j)|
+|D({\bf z}^*,{\bf a}_2^*;{\bf a}_1,{\bf a}_2)||D({\bf a}_1^*,{\bf b}_j^*;{\bf z},{\bf a}_j)|\Big\}. \eqno(4.28)$$
We can confirm that $|D|^2P_3=0$ by virtue of Jacobi's identity (3.7) and the formula (3.8) with $n=3$. Thus, we complete the proof of Eq. (2.6). \par
\bigskip
\leftline{\bf 5. Concluding remarks}\par
\leftline{\it 5.1. Generalization to the $n$-component system}\par
The generalization of the system (1.1)  to the $n$-component system $(n\geq 3)$  will be obvious. It takes the form
$${\rm i}\,q_{j,t}+q_{j,xx}+\mu\left(\sum_{k=1}^n|q_k|^2\right)q_j+{\rm i}\gamma\left[\left(\sum_{k=1}^n|q_k|^2\right)q_j\right]_x=0,\quad (j=1, 2, ..., n). \eqno(5.1)$$
By means of the gauge transformation similar to (2.1), one can transform (5.1) to the system  (2.2) in which the summation with respect to $k$ extends from 1 to $n$. 
The system of bilinear equations corresponding to (2.4), (2.5) and (2.6) can be derived  easily. They read
$$({\rm i}D_t+D_x^2)g_j\cdot f=0, \quad (j=1, 2, ..., n), \eqno(5.2)$$
$$D_xf\cdot f^*={{\rm i}\gamma\over 2}\sum_{k=1}^n|g_k|^2, \eqno(5.3)$$
$$D_x^2f\cdot f^*=\mu\sum_{k=1}^n|g_k|^2+{{\rm i}\gamma\over 2}\sum_{k=1}^nD_xg_k\cdot g_k^*. \eqno(5.4)$$
The $N$-soliton solution of the above system of bilinear equations would be given by (2.9)  with the element $c_{jk}$ in (2.10b) 
being replaced by $\sum_{s=1}^n\alpha_{sj}\alpha^*_{sk}$. The key formulas in performing the proof of the $N$-soliton solution are the formula (3.8) and  the generalization of the
formula (3.9)
$$|A+\sum_{s=1}^n{\bf b}_s^T{\bf a}_s|=|A|+\sum_{m=1}^{n^\prime}(-1)^m\sum_{1\leq s_1<s_2<...<s_m\leq n}
|A({\bf a}_{s_1},{\bf a}_{s_2},...,{\bf a}_{s_m} ;{\bf b}_{s_1},{\bf b}_{s_2},...,{\bf b}_{s_m})|, \eqno(5.5)$$
 where $n^\prime={\rm min}(n, M)$. 
 \par
 \medskip
 \leftline{\it 5.2. Nonlocal modified NLS equation}\par
Another interesting issue to be worth studying is a nonlocal coupled NLS and derivative NLS equation arising from a
continuum limit of the system (5.1), which is written as
$${\rm i}\,q_t+q_{xx}+\mu\left(\int_{-\infty}^\infty|q|^2dz\right)q+{\rm i}\gamma \left(\int_{-\infty}^\infty|q|^2dz\, q\right)_x=0, \quad q=q(x,t,z). \eqno(5.6)$$
When $\gamma=0$, this equation coincides with a (2+1)-dimensional nonlocal NLS equation proposed by Zakharov [8]. Its $N$-soliton solution has been obtained in [9].
The bilinear equations corresponding to (5.2), (5.3) and (5.4) stem from their continuum versions where $g_j$ is replaced by $g=g(x,t,z)$ and the sum with respect to $k$
chages to the integral with respect to $z$. We can expect that the $N$-soliton solution of Eq. (5.6) takes the same form as (2.9) with $c_{jk}$ in (2.10b) being replaced by
$\int_{-\infty}^\infty\alpha_j(z)\alpha_k^*(z)dz$ where $\alpha_j(z)\ (j=1,2,..., n)$ are continuous functions of $z$.
However, its proof still remains open.
The detailed analysis of the problems mentioned above will be done in future works. \par
\medskip
\leftline{\it 5.3. Miscellaneous remarks}\par
\noindent 1. The results in this Letter have wide applications in the description of the propagation characteristics of ultrashort pulses in an optical fiber.
One of which has been presented in [4] for investigating the dynamics of two optical solitons. The interaction process of an arbitrary number of
solitons would be worth studying. As for various applications, one can refer to the references [1], [3] and [4] as well as a number of papers cited therein. \par
\noindent 2.  The extended or  higher-order NLS equations  have 
been studied previously by means of the bilinear method. Unlike the system (1.1), they require the third-order dispersion.   On can consult the references [10] and [11], for example as for the construction of the
$N$-soliton solutions . \par
\noindent 3. An advantage of using the determinantal expressions of the solution is that the proof of bilinear equations reduce to the well-known formulas
for determinants, as demonstrated in this Letter. On the other hand, it is possible to represent the solution in the form of a finite sum of exponential 
functions. See, for example, an $N$-soliton solution of a single-component modified NLS equation  in which the
proof of the $N$-soliton solution has been performed by an ingenious mathematical induction [6]. Its proof is, however,  not easy to understand when compared with an elementary 
proof relying on Jacobi's identity. \par
\noindent 4. The repeated application of the Darboux transformation to the system (1.1) with $\mu=0$ yields  the bright $N$-soliton solution of the form
$q_1=(H_2/H_1)_x, q_2=(H_3/H_1)_x$ where $H_j (j=1,2, 3)$ are $3N\times 3N$ determinants [5]. In terms of the bilinear operator, they can be
rewritten as $q_1=D_xH_2\cdot H_1/H_1^2, q_2=D_xH_3\cdot H_1/H_1^2$. Comparing these solutions with the $N$-soliton solutions (2.9) with $\mu=0$, 
one can infer the relations $f=cH_1, f^*g_1=c^2D_xH_2\cdot H_1$ and $f^*g_2=c^2D_xH_3\cdot H_1$ with $c$ being a complex constant.
The proof of  these relations seems to be  a highly nontrivial problem to be pursued in a separate context.\par
\bigskip
\leftline{\bf Acknowledgement}\par
This work was partially supported by the Grant-in-Aid for Scientific Research (C) No. 22540228 from Japan Society for the Promotion of Science. \par
\bigskip
\leftline{\bf References}\par
\begin{enumerate}[{[1]}]
\item M. Hisakado, T. Iizuka and M. Wadati, J. Phys. Soc. Jpn. 63 (1994) 2887.
\item M. Hisakado and M. Wadati, J. Phys. Soc. Jpn.  64 (1995) 408.
\item A. Janutka, J. Phys. A  41 (2008) 285204.
\item H.Q. Zhang, B. Tian, X. L\"u, H. Li and X.H. Meng, Phys. Lett. A 373 (2009) 4315. 
\item L. Ling and Q.P. Liu, J. Phys. A 43 (2010) 434023.
\item S.L. Liu and W.Z. Wang, Phys. Rev. {\bf E48} (1993) 3054.
\item R. Vein and P. Dale, Determinants and Their Applications in Mathematical Physics, Springer, New York, 1999.
\item V.E. Zakharov, in: R.K. Bullough, P.J. Caudrey(Eds.), Solitons, in: Topics in Current Physics, Springer, Berlin, 1980, p. 243.
\item K. Maruno and Y. Ohta, Phys. Lett. A 372 (2008) 4446.
\item S.L. Liu and W.Z. Wang, Phys. Rev. {\bf E49} (1994) 5726.
\item C. Gilson, J. Hietarinta, J. Nimmo and Y. Ohta, Phys. Rev. {\bf E68} (2003) 016614.
\end{enumerate}

\end{document}